\documentclass[superscriptaddress, amsmath,amssymb, pre, twocolumn,floatfix]{revtex4-1}

\usepackage{graphicx}
\usepackage{dcolumn}
\usepackage{bm}
\usepackage{hyperref}
\usepackage{color}
\usepackage{amsmath}
\usepackage{amssymb}

\begin{document}

\title{Dynamical patterns in active nematics on a sphere}

\author{Silke Henkes}
\affiliation{Institute of Complex Systems and Mathematical Biology, Department of Physics, University of Aberdeen, Aberdeen AB24 3UE, United Kingdom}
\email{shenkes@abdn.ac.uk}
\author{M.~Cristina Marchetti}%
\affiliation{Department of Physics and Soft Matter Program, Syracuse University, Syracuse, NY 13244, USA}%
\email{mcmarche@syr.edu}
\author{Rastko Sknepnek}%
\affiliation{School of Sciences and Engineering and School of Life Sciences, University of Dundee, Dundee, DD1 4HN, United Kingdom}%
\email{r.sknepnek@dundee.ac.uk}

\date{\today}

\begin{abstract}
Using agent-based simulations of self-propelled particles subject to short-range repulsion and nematic alignment we explore the dynamical phases of a 
dense active material confined to the surface of a sphere. We map the dynamical phase diagram as a function of curvature, alignment strength 
and activity and reproduce phases seen in recent experiments on active microtubules moving on the surfaces of vesicles. At low driving, we recover the 
equilibrium nematic ground state with four $+1/2$ defects. As the driving is increased, geodesic forces drive the transition to a band of polar matter wrapping 
around an equator, with large bald spots corresponding to two $+1$ defects at the poles. Finally, bands fold onto themselves, followed by the  system moving 
into a turbulent state marked by active proliferation of pairs of topological defects. We highlight the role of nematic persistence length and time for pattern 
formation in these confined systems with finite curvature. 
\end{abstract}

\pacs{Valid PACS appear here}

\maketitle

\section{Introduction}

Active matter consists of self-driven agents that individually dissipate energy and organise in collective self-sustained 
motion. Active systems are maintained  out of equilibrium by a drive that acts independently on each agent, breaking  
time reversal symmetry \emph{locally}, rather than globally as in more familiar condensed matter systems driven out 
of equilibrium by external fields or boundary forces. Realisations span many scales, both in the living 
and non-living world, from bird flocks to bacterial suspensions, epithelial cell layers and synthetic 
microswimmers~\cite{marchetti2013hydrodynamics,ramaswamy2010mechanics}. 

Active particles are often elongated and form active liquid crystal phases, with either polar or nematic symmetry, and emergent patterns
controlled by the interplay of orientational order and active flows. Novel effects predicted and observed in simulations and experiments 
include  long-range order in two dimensions~\cite{vicsek1995novel}, spontaneous laminar flow~\cite{voituriez2005spontaneous}, giant number 
fluctuations~\cite{simha2002hydrodynamic,narayan2007long,chate2006simple}, novel rheology~\cite{hatwalne2004rheology,marchetti2012active}, 
and active turbulence accompanied by the proliferation of topological defects that drive the self-sustained dynamics~\cite{sanchez2012spontaneous}. 
Very recently, topological defects have also been shown to affect cell death and extrusion in nematically ordered epithelia~\cite{saw2017topological} 
and neural progenitor cell cultures~\cite{kawaguchi2017topological}. Much of this rich behaviour is captured well by active hydrodynamics.

Even more surprises arise when active systems are confined by bounding surfaces or to geometries that require defects in the orientational order. 
When confined to a box, active particles accumulate at the boundary, with the strongest accumulation at corners, demonstrating the dramatic effects 
of wall curvature~\cite{yang2014spiral,fily2016active}. When constrained to move on curved surfaces, as realised for instance when cells migrate in 
the gut epithelium~\cite{sato2009single} or on the surface of the growing cornea~\cite{collinson_clonal_2002}, the interplay of activity and curvature 
can drive novel dynamical structures. Keber \emph{et al.}~ \cite{keber2014topology} have studied this interplay under controlled conditions in active 
vesicles obtained by confining an active nematic suspension of kinesin-microtubule bundles to the surface of a lipid vesicle. Topological defects are 
unavoidable when a nematic liquid crystal is confined to the surface of a sphere, where the net topological charge must be $2$. In equilibrium the 
lowest energy configuration consists of four $+1/2$ disclinations arranged at the corner of a tetrahedron inscribed in the sphere~\cite{nelson2002toward,shin2008topological}. 
In the active vesicles of Ref.~\cite{keber2014topology}, this four-defect configuration oscillates at a well defined rate controlled by the concentration 
of ATP between the tetrahedral configuration and a planar one, with the four defects on an equator. A minimal model of $+1/2$ active defects as self-propelled 
particles~\cite{giomi2013defect}, subsequent hydrodynamic descriptions~\cite{khoromskaia2016vortex} and a particle-based extensile nematic 
simulation~\cite{alaimo2017curvature} reproduce the  oscillatory behaviour, but are inadequate to describe the rich succession 
of dynamical states observed when decreasing the size of the vesicles, including defect-driven  protrusions, and two $+1$ aster defects at the poles 
with spontaneously folding nematic  bands. 

\begin{figure*}[t]
\centering
\includegraphics[width=1.0\textwidth]{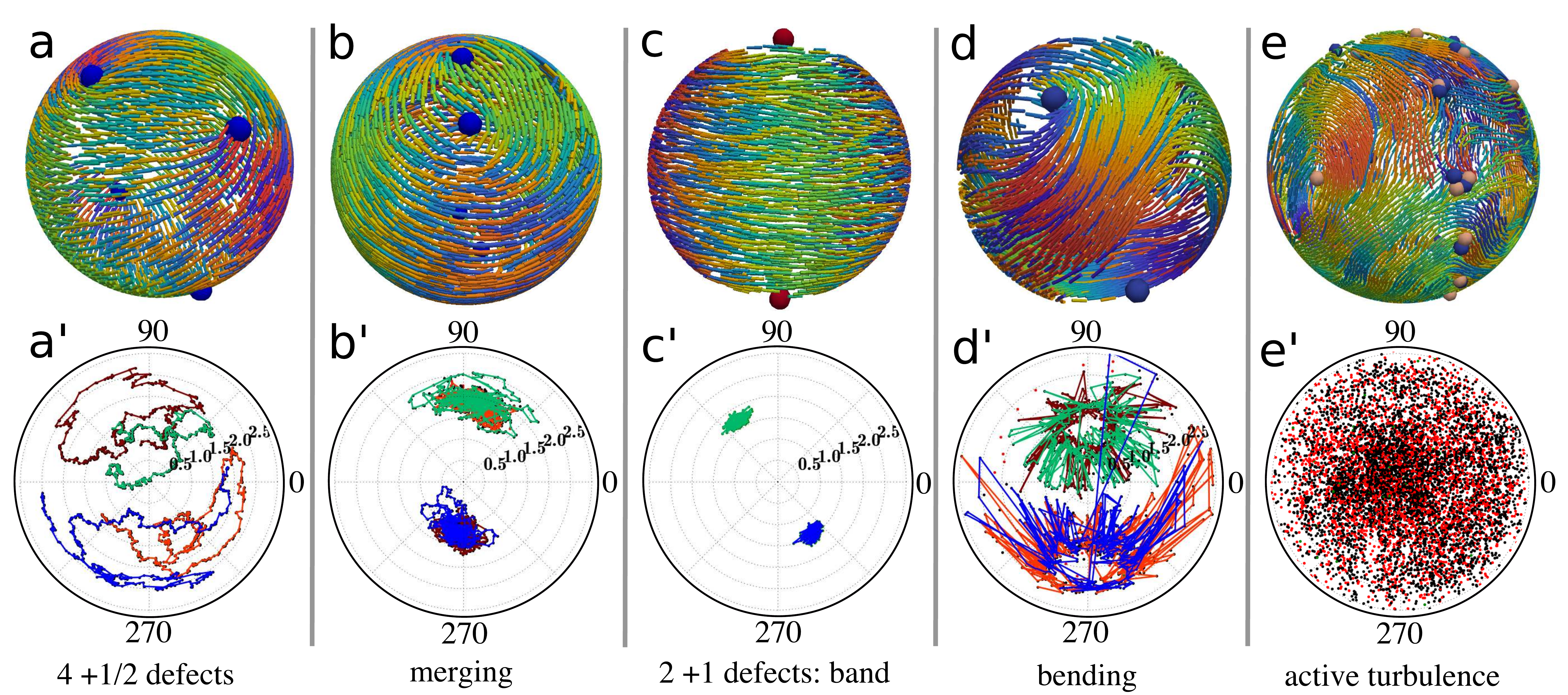}
\caption{Active nematic patterns formed by agents self-propelled at speed $v_0$ on a  sphere of radius $R$. Individual agents randomly reverse their direction 
of self propulsion at rate $\tau_{\text{flip}}^{-1}$ and align nematically with neighbours at rate $J$.  Activity (quantified by $v_0$) increases from left to right 
(see also supplementary movies M1-M5). Top row: Nematic patterns and defect positions. Individual agents are represented by bars, coloured by the x-component 
of their director. Defects are represented by dots, coloured by charge (dark blue: $+1/2$, tan: $-1/2$, red: $+1$).  Polarisation sorting and the formation of lanes 
traveling in opposite directions is evident in (d) and (e). Bottom row: Stereographic projections of defect trajectories in the corresponding states (see also Fig. S3 
for the corresponding tracks). Here $+1/2$ / $-\!1/2$ defects are represented by black/red dots, $+1$ defects are green dots, and where possible, we have tracked 
defect trajectories in different colours. All images are for $\tau_{\mathrm{flip}}=100$ and $R=30$, except for (e) where $R=50$. (a, a') Four $+1/2$ defects for $v_0=0.4$ 
and $J=1$; (b, b') merging defect state for $v_0=1$, $J=5$ (b) and  $v_0=1.2$ and $J=2$ (b'); (c, c') band state for $v_0=1$, $J=0.1$ (c) and  $v_0=1.2$ and $J=0.1$ (c'); 
(d, d') bending bands for $v_0=2$, $J=0.1$ (d) and $v_0=2.5$ and $J=0.1$ (d'); (e, e') active turbulence for  $v_0=4$ , $J=1$ (e) and $v_0=5$ and $J=5$ (e').}
\label{fig:snapshots}
\end{figure*}

In this paper, we examine the rich dynamics of active nematics on a sphere by considering a model of soft self-propelled active 
agents with nematic alignment. The model  reproduces a number of the curvature and activity induced dynamical structures observed in the experiments of 
Ref.~\cite{keber2014topology}, including the oscillating state of four $+1/2$ disclinations, an equatorial nematic band, a state where the band folds on itself, 
and eventually the transition to turbulence with proliferation on unbound defect pairs. This rich succession of states obtained with increasing activity is shown in 
Fig.~\ref{fig:snapshots}.
  
Circulating band states have also been reported by two of us in a simulation of soft self-propelled agents on a sphere, but with polar alignment~\cite{sknepnek2015active}. 
Recent work by one of us has additionally demonstrated that band formation on a sphere is a universal feature of the Toner-Tu equations for polar flocks ~\cite{shankar2017topological}. 
It was also recently shown that  on ellipsoids, the band localises to the low-curvature region \cite{ehrig2016curvature}. Here we show that active nematics also exhibit band 
states arising from the interplay of active motion and curvature. Our work shows that the formation of the band state is a generic property of active systems, independent 
of their symmetry and that it arises from the intrinsic tendency of active particles to move along geodesics on curved surfaces. The identification of this key mechanism 
for driving pattern formation on curved topologies is an important result of our work.

A second, more subtle, finding concerns the emergence of hydrodynamics from agent-based models. Derivations of hydrodynamic equations from microscopic dynamics 
can be carried out at low density~\cite{marchetti2013hydrodynamics}, but become challenging in the dense limit~\cite{gao2015multiscale}. In particular, although our system 
behaves like a nematic fluid at  long times and large length scales, it also exhibits a substantial amount of short range polar order and local flocking at intermediate time scales, 
suggesting that a suitable continuum model may require inclusion of a polarisation field in conjunction to a nematic order parameter. A similar result was reported for a related model 
in the plane~\cite{shi2013topological}, hence is not the result of curvature. This behaviour is also reminiscent of that of collections of self-propelled rods~\cite{bertin2015comparison} 
- polar active agents that exhibit nematic order at large scales. This local polar order becomes important in confined geometries like a sphere when the persistence length of the 
motion becomes comparable to the sphere radius and drives polarity sorting and leads to the folding band regime shown in Fig.~\ref{fig:Bending}. In the experiments of 
Ref.~\cite{keber2014topology} similar structures occur when the length of the microtubule bundles is comparable to the size of the vesicle. 

The paper is organised as follows. In Sec.~\ref{sec:model} we introduce the model of soft active agents moving on the  surface of a 2-sphere. In Sec.~\ref{sec:analysis}, we outline 
methods used to analyse behaviour of the system, followed by a detailed characterisation of the five distinct dynamical patterns that we have observed presented in Sec.~\ref{sec:results}.
Finally, we conclude in Sec.~\ref{sec:conclusions}. In Appendix~\ref{sec:appendix} we introduce the algorithms used for identifying and tracking defects.
 
\begin{figure}[t]
\begin{centering}
\includegraphics[width=0.9\columnwidth]{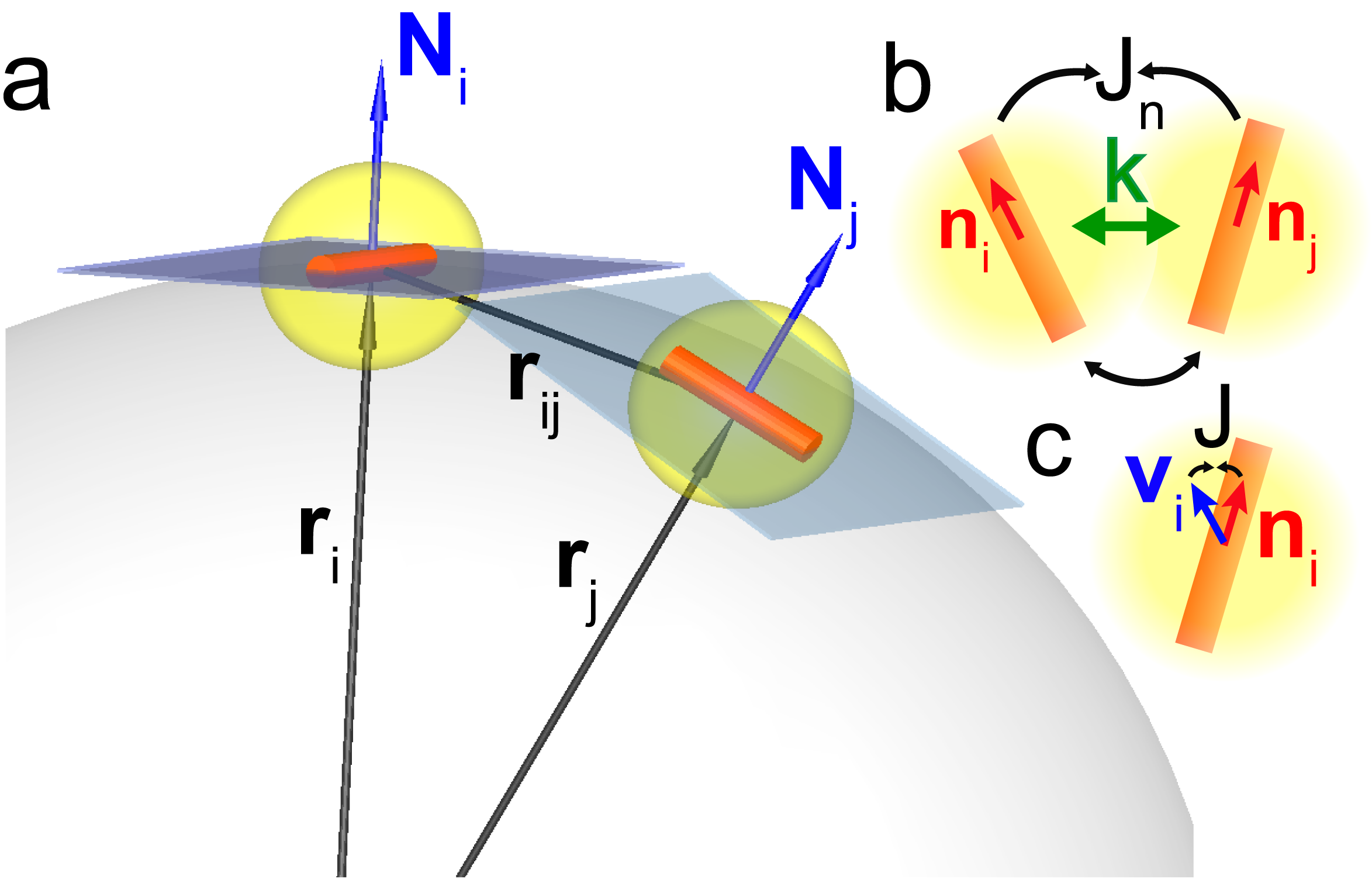}
\end{centering}
\caption{(Colour online)  (a) Self-propelled agents are constrained to the surface of a sphere of radius $R$. Agents have an internal 
degree of freedom, the director $\mathbf{n}_i$ (red/dark cylinders). The position of each agent is given by $\mathbf{r}_i$, measured 
from the centre of the sphere. Distances between agents, $r_{ij}=\left|\mathbf{r}_i-\mathbf{r}_j\right|$, are calculated as the Euclidean distance in $\mathbb{R}^3$ 
space. (b) Agents are subject to a soft-core repulsion and experience torque which nematically aligns them to their immediate neighbours. 
(c) Additionally, each agent has a preference to align its director to the direction of its motion. 
 \label{fig:model} }
\end{figure}

\section{Model: Soft active agents on a sphere}
\label{sec:model}
Active particles move through a medium that mediates long-range hydrodynamic interactions. Here we consider  
the \emph{dry} limit, where the damping from the medium dominates over hydrodynamic interactions and the fluid only provides single-particle friction. 
Our system consists of $N$ self-propelled soft spherical agents of radius $\sigma$ confined to move on the surface of a 2-sphere of radius $R$ (Fig.~\ref{fig:model} ).
Each agent is characterised by its position $\mathbf{r}_i$ and a unit vector $\mathbf{n}_i$ or director denoting the direction of self-propulsion.
Although confinement to the surface of a sphere implies that agent positions are parametrised by just two coordinates, e.g.
azimuthal and polar angles, in numerical simulations it is more convenient to work in three-dimensional Euclidean space and explicitly impose the sphere 
constraint at each time step. This is accomplished through projection operators. The coupled equations for translational and rotational motion are then given by
 
\begin{subequations}
\label{eq:equations_of_motion}
\begin{eqnarray}
\dot{\mathbf{r}}_{i} &= &\mathbf{P}^{T}_i\big[v_0\kappa_i(t)\mathbf{n}_{i}+\mu\sum \nolimits_{j}\mathbf{F}_{ij}\big]\;,\label{eq:motion_r} \\
\dot{\mathbf{n}}_i &=&\text{P}^N_i\big[- 2 J_n \sum \nolimits_j  (\mathbf{n}_i\cdot\mathbf{n}_j)(\mathbf{n}_i\times  \mathbf{n}_j) \nonumber \\
                             &-&2J_v (\mathbf{v}_i\cdot\mathbf{n}_i)(\mathbf{v}_i\times\mathbf{n}_i) \big] (\hat{\mathbf{N}}_i \times  \mathbf{n}_i)\;, \label{eq:motion_n}
\end{eqnarray}
\end{subequations}
where  $\mathbf{P}^{T}_i[\mathbf{a}]=\mathbf{a}-(\hat{\mathbf{N}}_i\cdot\mathbf{a})\hat{\mathbf{N}}_i$ and $\text{P}^N_i[\mathbf{a}]=(\mathbf{a} \cdot \hat{\mathbf{N}}_i)$ 
project any three-dimensional vector $\mathbf{a}$ onto the tangent plane and the surface normal to the sphere, respectively, with the local surface normal given 
by $\mathbf{\hat{N}}_i=\mathbf{r}_i/r_i$~\cite{sknepnek2015active}. 
 
The first term on the right hand side of Eq.~\eqref{eq:motion_r} describes self-propulsion at speed $v_0$. The nematic symmetry is implemented by 
 reversing the direction of  self propulsion at time intervals drawn from a Poisson distribution of mean $1/ \tau_{\text{flip}}$ through a zero average bimodal noise 
 $\kappa_i(t)=\pm 1$ \cite{chate2006simple}. The second term in Eq.~\eqref{eq:motion_r} describes soft pairwise repulsive forces of stiffness $k$,
$\mathbf{F}_{ij}=-k\left(2\sigma-r_{ij}\right)\frac{\mathbf{r}_{ij}}{r_{ij}}$, with $\mathbf{r}_{ij}= \mathbf{r}_j-\mathbf{r}_i$ and $r_{ij} =|\mathbf{r}_{ij}|$, for $r_{ij}<2\sigma$ 
and $\mathbf{F}_{ij}=0$ otherwise  (Fig.~\ref{fig:model}b), and $\mu$ the mobility. Here $r_{ij}$ is the Euclidean distance computed in $\mathbb{R}^{3}$. The rotational 
dynamics of $\mathbf{n}_i$ is governed by Eq.~\eqref{eq:motion_n}. It contains two contributions to the torque: (i) nematic alignment of neighbouring directors at 
rate $J_n$, with the sum extending over all neighbours within a radius $r_c=2.4\sigma$; and (ii) alignment of the director $\mathbf{n}_i$ with the direction of its own 
velocity, $\mathbf{v}_i=\dot{\mathbf{r}}_i$, at rate $J_v$.  
 
In the absence of interactions, the mean-square  displacement of a single agent is diffusive at long times with effective diffusion constant $D_0=v_0^2\tau_{\text{flip}}/4$. 
The random flips serve the same function as rotational noise in Active Brownian Particle models~\cite{cates2015motility} and we therefore neglect rotational noise in 
Eq.~\eqref{eq:motion_n}. Additionally we neglect Brownian noise in the translational dynamics because we focus on the high density regime of a sphere with packing 
fraction $\phi=N\sigma^2/4R^2=1$, where the dynamics is dominated by steric repulsion and noise is negligible compared to the effects of collisions. 

 To make contact with more familiar forms of the equations of motion for active nematic agents, we note that in the local tangent plane at position $\mathbf{r}_i$ we can 
 write $\mathbf{n}_i=\left(\cos\theta_i,\sin\theta_i\right)$, where $\theta_i$ is the angle with the local $x-$axis. Then the $J_n$-term yields a torque proportional to 
 $\sin2(\theta_j-\theta_i)$ (up to correction terms of order $2\sigma/R$ due to parallel transport), which is the nematic coupling used in previous literature~\cite{chate2006simple}. 
In the same coordinate system the second term on the right-hand-side in Eq.~\eqref{eq:motion_n} takes the form $J_v \sin2(\theta^v_i-\theta_i)$, where $\theta^v_i$ is 
the angle of the velocity vector with the local $x$ axis in the tangent plane, hence it tends to align the director with the particle's velocity~\cite{szabo2006phase,henkes2011active}. 
This term is required to ensure torque transfer between orientational and translational degrees of freedom, since for spherical agents changes in the direction of motion are 
not directly coupled to the director $\mathbf{n}_i$. It is needed to drive the nucleation of defect pairs. In models with anisotropic agents, such coupling arises naturally from 
steric interactions that are anisotropic and transfer torque, ``stirring''  the local structure when an agent changes its direction. For simplicity, in the following we set 
$J_{v}=J_{n}\equiv J$.
 
To date, three other microscopic models of active nematics that fully include steric effects have been investigated with large-scale simulations. In planar geometry, 
DeCamp, \emph{et al.}~\cite{decamp2015orientational} simulated a collection of elongated rod-like particles whose length grows at a steady rate, directly producing an extensile stress, 
until at a critical length a rod splits into two while two other rods merge to maintain number conservation. This model exhibits an active nematic state with somewhat different 
defect symmetries than observed in the experiment of Keber, \emph{et al.}~\cite{keber2014topology}. Shi and Ma \cite{shi2013topological} studied a model of self-propelled
ellipsoids with reversal  of the direction of self propulsion similar to the one employed in the present study. They focused on the regime of very short
 $\tau_{\mathrm{flip}}$ and found very weakly extensile forces in this region of parameters, together with very slowly moving $+1/2$ defects. 
As we will show below, our results confirm the observations of Shi and Ma that defects are only very weakly motile and the dynamics is dominated by strongly fluctuating locally polar flows (see Fig. S2).  
Very recent work by Alaimo \emph{et al.}~\cite{alaimo2017curvature} on spheres and ellipsoids introduces a model with the same steric components as ours, but with a directly 
extensile forcing instead of self-propulsion. This leads to a locally extensile material with an oscillating defect phase, but in contrast no band phase was reported. 

Here we focus on reversing self-propelled systems where reversal is slow compared to the particle collision time, i.e., $\tau_{\text{flip}}\gg(\mu k)^{-1}$, where $(\mu k)^{-1}$ 
is the time scale of the interaction. This choice results in persistent dynamics on length scales  $\xi = v_0 \tau_{\text{flip}}$ that at large activity $v_0$ can become comparable to the size of the sphere. This is indeed the regime relevant to the experiments of Keber \emph{et al.}~\cite{keber2014topology}, where kinesin 
motors ``walk'' persistently along  the microtubules that have lengths comparable to the radius of the vesicles. Finally, in the following we measure time in units of  $\mu k$ 
and lengths in units of $\sigma$ and set $\tau_{\mathrm{flip}}=100$ (except where stated otherwise). 
 
\begin{figure}[h]
\centering
\includegraphics[width=0.99\columnwidth]{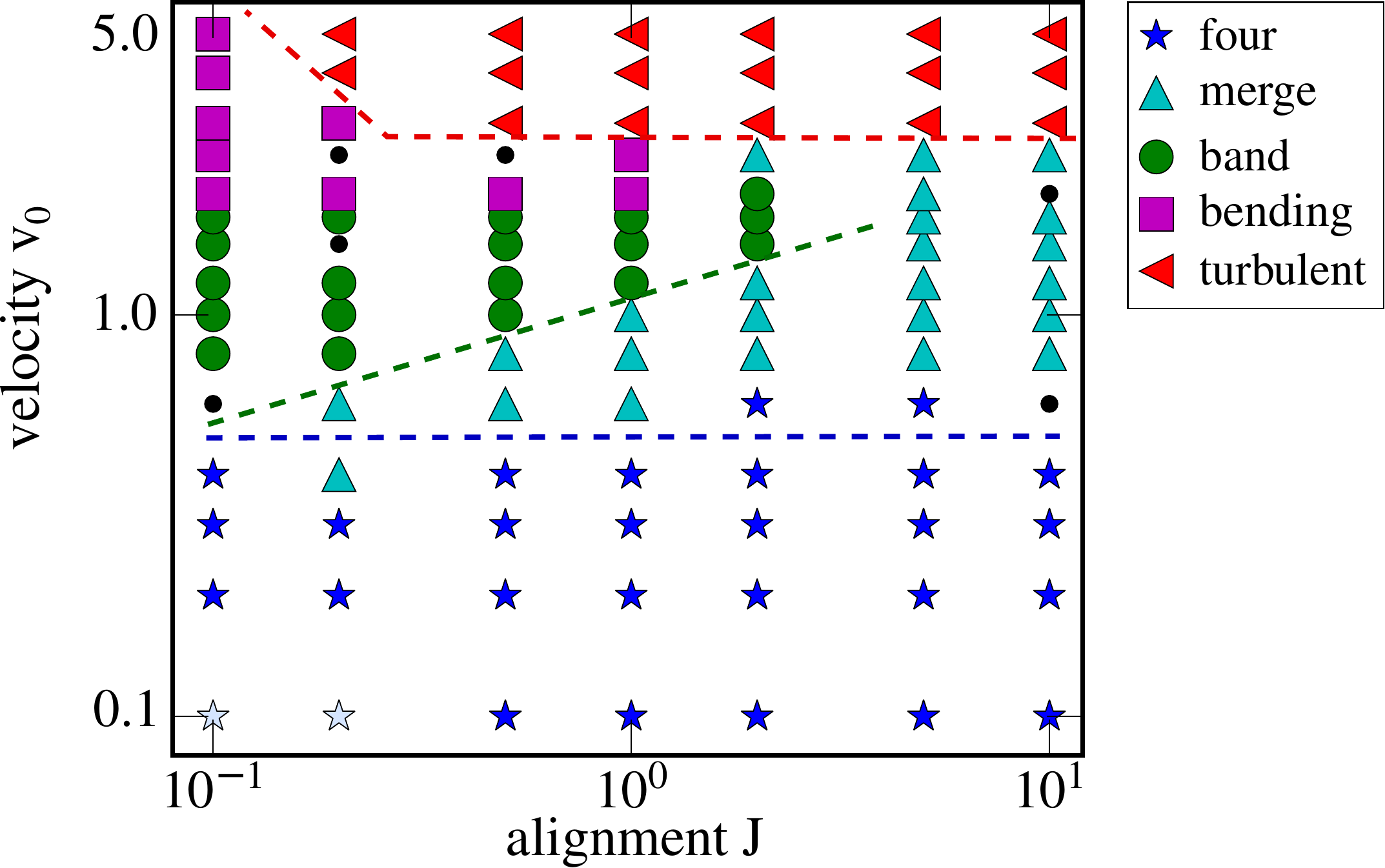}
\caption{Phase diagram at high $\tau_{\mathrm{flip}}=100$ as a function of coupling $J$ and of active driving $v_0$. We observe a low $v_0$ 
phase with four $+1/2$ defects (blue stars, pale blue stars) which gives way to an intermediate set of band states with either merging defects (turquoise upright triangles) 
or $+1/2$ defects fully merged into two $+1$ defects at low $J$ (green circles). This phase is succeeded by turbulence above a threshold $v_0$ (red sideways triangles). 
At low $J$, a folding instability in the band shapes appears as driving increases (purple squares). Some states defied classification (black dots). The coloured lines are guides 
to the eye representing the phase boundaries. }
\label{fig:phase_diagram}
\end{figure}

\section{Characterising dynamical patterns and defect structures}
\label{sec:analysis}
In this section we briefly outline the methods used to analyse the results of our Brownian dynamics simulations. 
A summary of our method for identifying and tracking defects using a tessellation method in the presence of density fluctuations is given in Appendix~\ref{sec:appendix}.
\subsection{Mean-square displacements} 
\vspace{-4mm}
To compute the mean squared displacement (MSD) of the defects, we used geodesic distances, i.e. the minimal distance along the great circle connecting the two positions in question.
For two arbitrary points $\mathbf{r}_i$ and $\mathbf{r}_j$ on the $2-$sphere, we have
\begin{equation} 
d_{ij} = R \arccos \left(\frac{\mathbf{r}_i}{R} \cdot \frac{\mathbf{r}_j}{R} \right).
\end{equation}
Then the MSD is computed as
\begin{equation}
 \text{MSD}(\delta t) = \langle d(t+\delta t) d(t) \rangle_{t,\text{defects}}. 
 \end{equation}
Tracking was not always successful, especially at larger $v_0$, and we excluded trajectories with too many outliers from the MSD computations. For smaller numbers of outliers, we employed 
an error-correction algorithm that interpolates the outlier position from the previous and next correctly tracked points. The scaling of the MSD as MSD $\sim t^{\alpha}$ allowed us to identify 
diffusive ($\alpha=1$) and persistent $(\alpha>1$) regimes, and to compute the diffusion coefficients of defect motion (Fig.~\ref{fig:MSD}).
\vspace{-2mm}
\subsection{Pair angles} 
\vspace{-4mm}
We define the pair angle between defects located at $\mathbf{r}_i$ and $\mathbf{r}_j$ as
\begin{equation} 
\theta_{ij} = \arccos \left(\frac{\mathbf{r}_i}{R} \cdot \frac{\mathbf{r}_j}{R} \right). 
\end{equation}
For the tetrahedron the pair angles are all $\theta_{ij}=109.47^{\circ}$, while in an alternative flat configuration seen in Ref.~\cite{keber2014topology}, four $90^{\circ}$ defects and two $180^{\circ}$ 
defects emerge, giving a mean pair angle $\langle \theta_{ij} \rangle =120^{\circ}$. We used $\langle \theta_{ij} \rangle$ to identify oscillations in Fig.~\ref{fig:oscillations}. The $+1$ defects of 
the band state are nearly separated by $180^{\circ}$, and in the merging state, pairs of $+1/2$ defects approach each other. Pair defect angle distributions are shown in Fig.~\ref{fig:Histograms}a, c. 
\subsection{Angular profiles} 
In the band state, a density profile emerges with bald spots at the poles and greater density at the equator. In parallel, the director field shows a systematic inclination
towards the equator. Both effects are due to curvature-induced forces~\cite{sknepnek2015active}. We reoriented configurations with the pole axis along $\mathbf{z}$ as follows. We first computed 
an estimate $\mathbf{z}_{\text{Ine}}$ of the axis $\mathbf{z}$ from the eigenvector associated to the smallest eigenvalue of the inertial tensor $\hat{I} = \sum_i \mathbf{r}_i \times \mathbf{r}_i$. 
Then we determined a $\pm 1$ orientation for each velocity vector $\mathbf{v}_i$ on the basis that $\mathbf{r}_i \times \mathbf{v}_i$ is largely either parallel or antiparallel to $\mathbf{z}_{\text{Ine}}$, 
and finally computed 
\begin{equation} 
\mathbf{z} = \sum_i \mathbf{r}_i \times \text{orient}(\mathbf{v}_i)  
\end{equation}
and normalised it. The density profiles in Fig.~\ref{fig:Histograms}b, d and the orientation profile of the director $\alpha_i = \arccos(\mathbf{n}_i \cdot e_{\phi})$ in Fig.~\ref{fig:Histograms}a 
have been averaged over all snapshots of several runs.

\section{Results}
\label{sec:results}

We have studied spherical vesicles of radii $R=5-50$ for packing fraction $\phi=1$, corresponding to full coverage of the sphere and $10^2-10^4$ agents. Results reported here are for $R=30$, unless stated otherwise. All our simulations are in a steady-state, 
in the sense that they have relaxed from an initial random configuration.  We study both the fast dynamics that immediately follows the relaxation to the steady state by integrating 
Eqs.~\eqref{eq:equations_of_motion} for a total of $1.5\times10^3$ time units with time step $\delta t=10^{-3}$ using a standard Euler algorithm, and the dynamics at long times with
 $\delta t=0.05$ for a total integration time of up to $10^{5}$ time units.

In Fig.~\ref{fig:snapshots}, we show the succession of states obtained with increasing activity $v_0$. As required by the Poincar\'{e}-Hopf theorem \cite{frankel2011geometry}, a nematic field
on the surface of a sphere always contains topological defects with total charge $2$. This can be satisfied by two $+1$ defects at the poles of the sphere, or four $+1/2$ defects
at the corners of a tetrahedron, or else by a turbulent state with many defects of both positive and negative charges, but adding up to a net charge of $+2$. The tetrahedral arrangement 
is the ground state configuration for a nematic with equal bend and splay Frank constants~\cite{nelson2002toward,shin2008topological}. In the presence of activity the rules governing 
conservation of topological charge are of course unchanged, but defects often become dynamical entities that move with actively driven flows. At low activity $v_0$ we obtained a texture 
of four well-separated $+1/2$ disclinations, as shown in Fig.~\ref{fig:snapshots}a. The four defects, however, are not  static, but move either diffusively or, at long times, in an oscillatory fashion.
At larger values of activity, pairs of $+1/2$ defects are pushed towards opposite poles (Fig.~\ref{fig:snapshots}b) and eventually merge resulting in a configuration of two
 $+1$ defects at the poles, with a band of nematic wrapping around a great circle  (Fig.~\ref{fig:snapshots}c). Agents move within the band, with approximately equal
fractions moving in clockwise and counterclockwise directions, so that nematic order is maintained with no mean flow. The ``bald spots'' at the poles can be interpreted as the cores of very large defects.
The width of the band increases when lowering the alignment coupling $J$, as observed in polar systems~\cite{sknepnek2015active}. At even higher activity and sufficiently low alignment $J$, 
the band becomes unstable to bend deformations (Fig.~\ref{fig:snapshots}d and Fig.~\ref{fig:Bending}) and one observes a complex folding dynamics, associated with sorting of particles 
into unidirectional lanes. Finally, upon further increase of $v_0$ the system exhibits turbulent-like flows with proliferation of topological defects (Fig.~\ref{fig:snapshots}e), akin to that 
observed in planar systems. 

We proceed to classify the various dynamical states depicted in Fig.~\ref{fig:snapshots} using the tools and methods discussed in previous section and the defect tracking algorithm 
outlined in Appendix~\ref{sec:appendix}. By combining all of these results, we are able to map the complex dynamical states observed in the simulations in the phase diagram shown 
in Fig.~\ref{fig:phase_diagram}.

\begin{figure}[h]
\centering
\includegraphics[width=0.99\columnwidth]{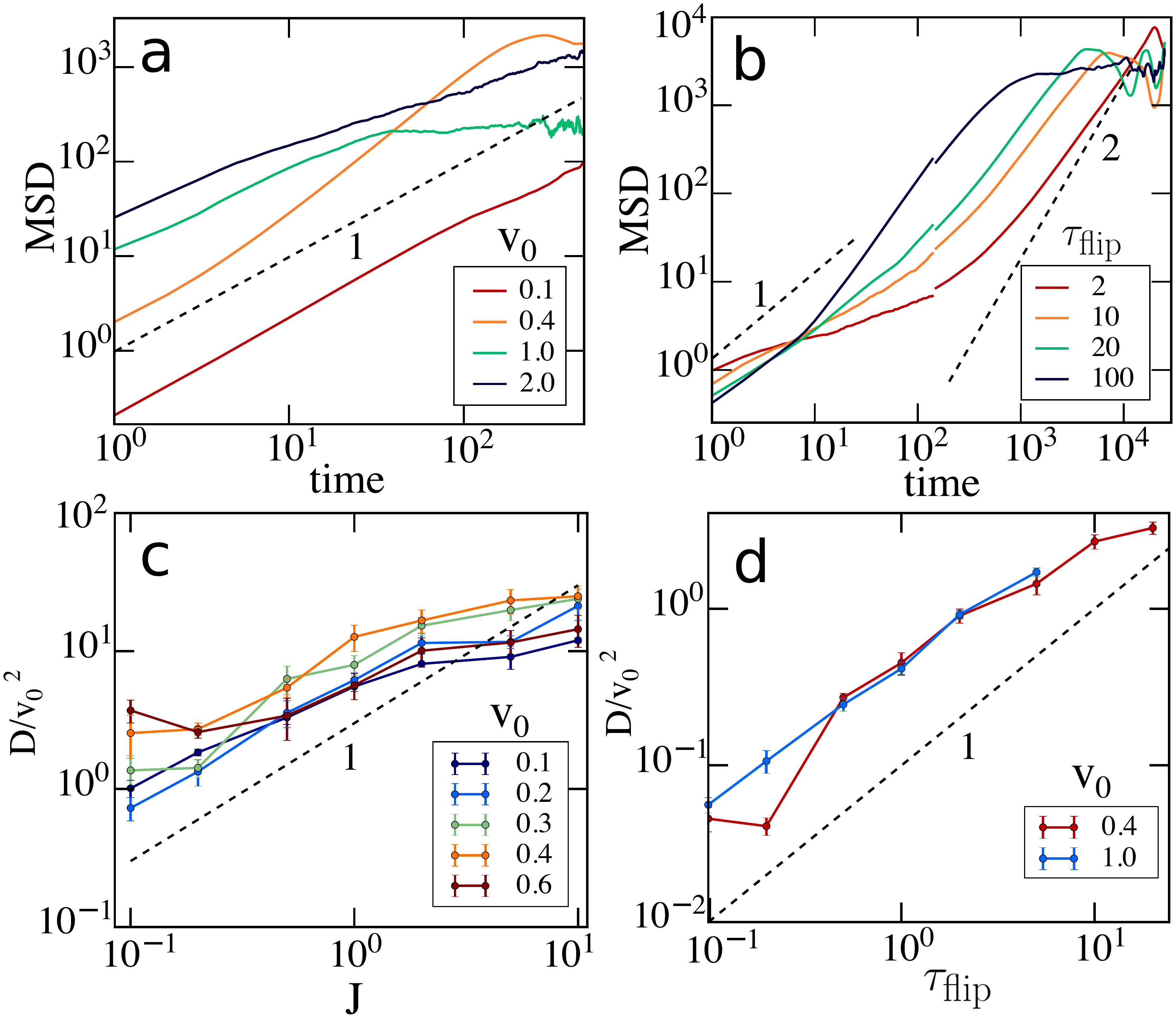}
\caption{Mean square displacements (MSDs) of defects, and diffusion coefficients. (a) Four representative MSD trajectories of $+1/2$ defects as a function of $v_0$ for $J=1.0$ 
and $\tau_{\mathrm{flip}}=100$. From bottom to top: diffusive motion at low $v_0$ becomes superdiffusive in regions with oscillations, then subdiffusive in the merging 
state, and close to diffusive again in the folding state. (b) The long-time MSD for oscillating systems at $J=0.1$ and $v_0=0.4$ becomes nearly persistent at different $\tau_{\text{flip}}$.
(c) Diffusion coefficient $D$ obtained from fits to the MSD, as a function of $J$ and scaled by $v_0^2$ for $\tau_{\mathrm{flip}}=100$. (d)  Scaled diffusion coefficient $D/v_0^2$ 
as a function of $\tau_{\mathrm{flip}}$. }
\label{fig:MSD}
\end{figure}

\subsection{Four-defect state}
\label{subsec:four-def}

The four defects are generally motile on the sphere, as shown in Fig.~\ref{fig:snapshots}  that displays the stereographic projections of their trajectories, with the only exception of the region 
of very low $v_0$ and $J$ where the system is effectively jammed (pale blue stars in the phase diagram Fig.~\ref{fig:phase_diagram} in a state
akin to the one reported by Fily, \emph{et al.}~\cite{fily2014freezing} and by Janssen, \emph{et al.}~on the sphere~\cite{janssen2016aging}). 
To analyse the defect configuration and their dynamics we have examined the histogram of the angles $\theta_{ij}$ of defect pairs (Fig.~\ref{fig:Histograms}a) and their 
mean-square-displacements (Fig.~\ref{fig:MSD}). We uncover two distinct regimes. At intermediate times, but long after settling into a long-lived steady state, the pair angle distribution is 
unimodal and peaked around $\theta=109^\circ$ corresponding to a tetrahedral configuration. In this regime, the MSD of the defects is diffusive (Fig. ~\ref{fig:MSD}a) with a diffusion coefficient 
that scales as $D\sim v_0^2\tau_{\text{flip}}$ as expected for particles self-propelled at speed $v_0$ with noisy flips of their direction of self-propulsion at rate $\tau_{\text{flip}}$. The diffusion rate 
also grows linearly with the alignment rate $J$, suggesting a scaling $D=v_0^2\tau_{\text{flip}}f(J\mu k)$, with $f(x)\sim x$ at small $x$. We note that in this regime the local flow and pressure fields 
are dominated by fluctuations (see Fig.~S2 in \cite{SI}), and we conjecture that the diffusive defect dynamics arises because defects are carried along by the strongly fluctuating flows.   
With increasing activity the distribution of pair angle broadens, but remains unimodal, and the MSD begins to cross over to superdiffusive behaviour at long times. Similar superdiffusive scaling 
was obtained by Shi and Ma for tapered rods \cite{shi2013topological} in the plane. 

\begin{figure}[t]
\centering
\includegraphics[width=0.95\columnwidth]{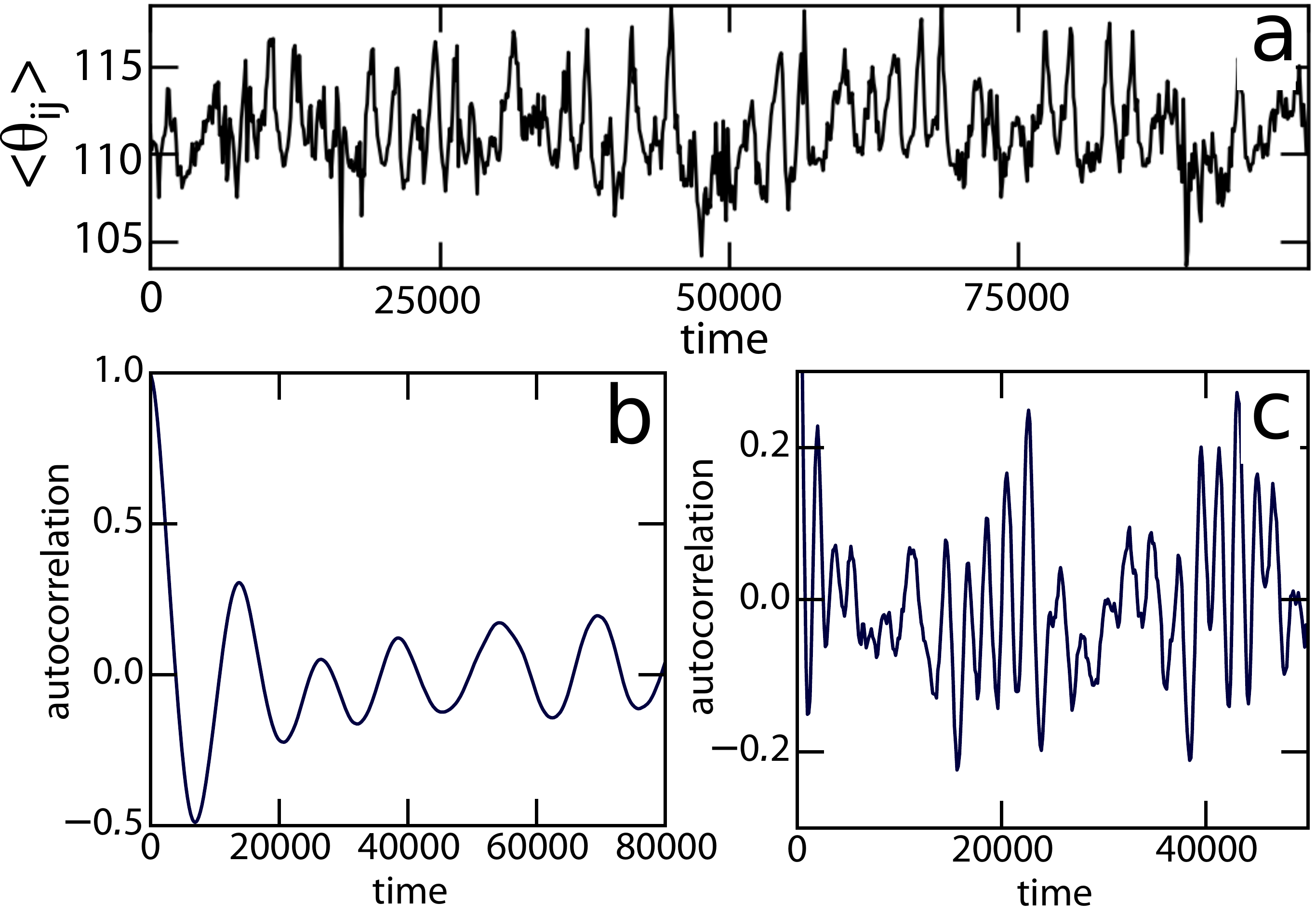}
\caption{Oscillations of the defects at long times, for $v_0=1.0$, $J=0.1$ and $\tau_{\text{flip}}=1.0$:
(a) mean pair angle $\langle \theta_{ij} \rangle$;  (b) autocorrelation function of individual defect trajectories, showing a single oscillation period of $T\approx10000$; (c) autocorrelation
 function of the mean pair angle, showing a superposition of several frequencies, with the fastest mode characterised by a period $T_m\approx 2000$.}
\label{fig:oscillations}
\end{figure}

\begin{figure}[t]
\centering
\includegraphics[width=0.99\columnwidth]{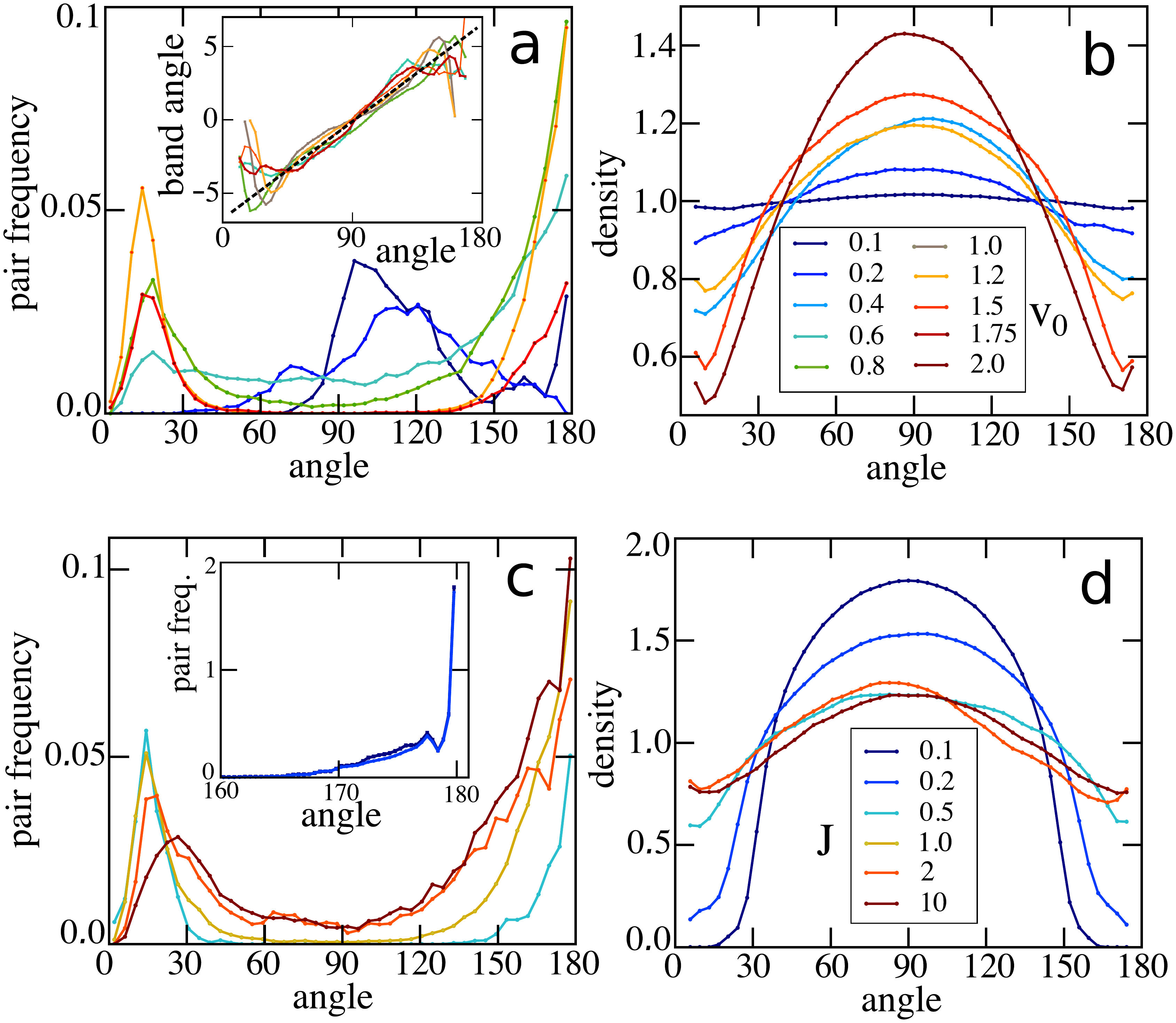}
\caption{Emergence of the band state as seen from the distribution of defect pair angles (a-c) and the  density profile (b-d). (a) At $J=1$, with increasing $v_0$,  the pair angle distribution 
evolves form unimodal with a single central peak ($v_0=0.1$) to bimodal with growing peaks close to $0^\circ$ and $180^\circ$ ($v_0=0.6-1.5$). The inset shows the inclination of the 
direction of self-propulsion given by the director field $\mathbf{n}$ relative to the equator as a function of latitude. The black dashed line has a slope of $0.07$.
(b) The corresponding density profiles evolve from homogeneous to banded.  (c) For $J=0.5-10$, at $v_0=1.0$, the distribution peaks of the merging state shift inwards. 
Inset: For $J=0.1-0.2$, the two $+1$ defects are nearly at opposite poles. (d) In the corresponding  density profile the bald spots at the poles disappear.}   
\label{fig:Histograms}
\end{figure}

At very long times the defect dynamics becomes superdiffusive with a largest exponent of approximately $\alpha\approx 1.5$,  and we observe oscillations akin to those reported in the experiments of 
Keber,\emph{et al.}~\cite{keber2014topology}, albeit only for specific parameter combinations. In Fig.~\ref{fig:MSD}b, we show the MSD trajectories over the full simulation time range in such a 
regime (corresponding to Fig.~S4 of \cite{SI}). We analyse the oscillations in Fig.~\ref{fig:oscillations} (see also supplementary movie M1). The mean pair defect angle (panel a) oscillates between 
approximately $109^\circ$ and $115^\circ$ and the oscillations are less regular than those reported in the experiment of Keber \emph{et al.} and in analytical \cite{khoromskaia2016vortex}
and numerical \cite{alaimo2017curvature} treatments, where the mean angle oscillates between $109^\circ$ and $120^\circ$, clearly reflecting oscillations between tetrahedral and planar 
defect configurations. To quantify the angular periodicity we have evaluated  the autocorrelation function of individual defect trajectories (panel b) and of the mean pair angle (panel c).   
Both display clear oscillatory behaviour, but the individual defect dynamics has a single oscillation period of $T\approx 20000$, while the mean angular dynamics 
is faster and controlled by several superimposed modes, with the fastest one oscillating at $T_m \approx 2000$. For $v_0=1.0$, $T$ increases with $\tau_{\mathrm{flip}}$ in the range sandwiched 
between frozen and band states. In Fig.~S4 in \cite{SI}, we show a corresponding analysis for $v=0.4$, where oscillations appear at much higher values of $\tau_{\mathrm{flip}} =10-50$. 
It is worth noting that we did not observe oscillations at any $J>0.1$, and we speculate that for larger $J$ the relaxation time $1/J$ of the director field is too fast for the flow field to coherently 
respond to it, so that again fluctuations dominate the dynamics.

\subsection{Band formation and merging defects}

As $v_0$ increases,we observe the emergence of  a nematic  band wrapped around an equator that is chosen through spontaneous symmetry breaking (see Fig.~\ref{fig:snapshots}b,c). The four 
$+1/2$ defects are pushed towards the poles, where they either form two nearly stationary trapped defect pairs (Fig.~\ref{fig:snapshots}b' and Fig. S3b) or merge into two fully stationary $+1$ defects 
(Fig.~\ref{fig:snapshots}c' and Fig.~S3c). This configuration has also been seen in the experiments of Ref.~\cite{keber2014topology}. The band state strongly resembles the band found in polar systems~\cite{sknepnek2015active}. It occurs because active particles are driven by curvature to move along geodesics, corresponding to great circles on a sphere. This effect is counterbalanced by repulsive interactions, resulting in the emergence of a finite-width band. Recent work by one of us \cite{shankar2017topological} has also shown that the spontaneous emergence of bands wrapping around the equator is a generic properties of active polar fluids that arises from the interplay spontaneous flow and curvature. In our nematic system the band forms only for sufficiently long  $\tau_{\text{flip}}$ ($\tau_{\mathrm{flip}}\geq 50$) when the 
system can support long-lived local polar flows that break the symmetry. Note that such bands are indeed seen in microtubule suspensions, where the defect structure shows clear nematic symmetry.
In Fig.~\ref{fig:Histograms} we examine the emergence of the band by looking at the evolution of the defect pair angle distribution (Fig.~\ref{fig:Histograms}a) and the density profile (Fig.~\ref{fig:Histograms}b) with increasing $v_0$.  The distribution of pair angles evolves from unimodal for $v_0=0.1-0.2$ corresponding to a fluctuating tetrahedral arrangement of four defects smeared-out by increasing 
diffusive motion to a  bimodal one with peaks close to the location of opposing poles when the defects merge and the band state emerges. Meanwhile the density profile (Fig.~\ref{fig:Histograms}b)
evolves from practically uniform to a distinct peak that continuously grows with $v_0$, because the curvature-induced force that drives band formation  is proportional to $v_0$~\cite{sknepnek2015active}. 
A similar behaviour is obtained by decreasing $J$ for fixed $v_0$ (Fig.~\ref{fig:Histograms}c,d). For low $J=0.1-0.2$, we always observe a band state with two $+1$ defects (inset to Fig.~\ref{fig:Histograms}c), 
and a very peaked density profile appears, with bald spots at the poles. At higher $J$, this gradually gives way to a nematic band with unmerged pairs of defects at the poles and the first peak position 
gradually shifts to larger angles and the whole distribution broadens (Fig. \ref{fig:Histograms}c).  We can explain this in part by noting that the defect core energy scales $\sim J$ when approximated by 
a single-constant Frank free energy \cite{chaikin2000principles} and so merging two $+1/2$ defects to form a $+1$ costs more energy at higher $J$. We see only slightly peaked density profiles for 
$J\geq 0.5$ (Fig.~\ref{fig:Histograms}d). The tendency of active agents to move along geodesics is confirmed by examining the angle between the direction $\mathbf{n}$ of self-propulsion relative to the 
equator~\cite{sknepnek2015active}. The angle profile across a set of bands as a function of latitude is shown in the inset of Fig.~\ref{fig:Histograms}c. The slope $\approx 0.07$ for the growth of the inclination
of $\mathbf{n}$ with latitude is much shallower than in the polar case for the same $J$, consistent with the suppressed appearance of nematic bands as compared to polar ones at high $v_0$.
Finally, in \cite{sknepnek2015active} we derived an effective profile for the polar band density. We show in \cite{SI} that this also provides a good fit for the density profile of nematic bands.

\begin{figure}[h]
\centering
\includegraphics[width=0.99\columnwidth]{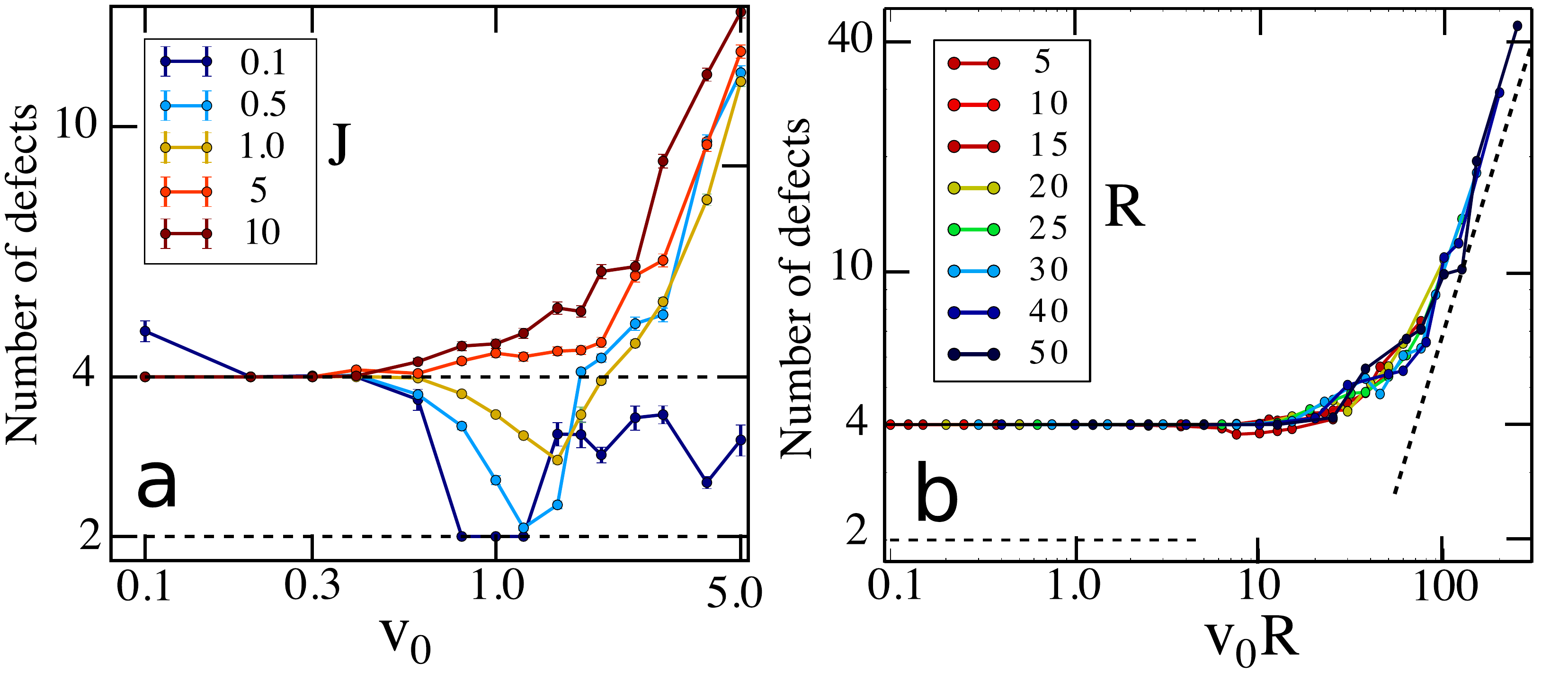}
\caption{(a) Number of defects of any charge as a function of $v_0$ and $J$ for $R=30$. At low $v_0$, all curves converge to the equilibrium ground state with
four $+1/2$ defects. As the driving is increased, the low $J$ systems develop a band state with two $+1$ defects. At the highest driving, the systems transition to a turbulent nematic
 state; lower $J$ delays onset of turbulence. (b) Defect scaling as a function of sphere radius for $J=10$. The onset of turbulence scales as $v_0 R$. The dashed diagonal line has a slope of 2.} 
\label{fig:Ndefects}
\end{figure}

\subsection{Transition to turbulence}

As even larger activity $v_0$, the system transitions to an active turbulent state, where pairs of $\pm 1/2$ defects are spontaneously created (Fig.~\ref{fig:snapshots}e' and Fig. S3e).
In Fig.~\ref{fig:Ndefects} we show the mean number of defects $N$ on the sphere regardless of their  charges. We identify $N>4$ as the onset of turbulence.

The turbulent  regime displays strong patterns of polar flow, consisting of lanes of particles moving in the same direction, as evident from Fig.~\ref{fig:snapshots}e. 
We estimate that in our confined system the transition to turbulence occurs when the spacing $l_{\alpha}$ between
defects becomes comparable to the sphere radius $R$. In hydrodynamic models in two spatial dimensions, the spacing between defects in the turbulent regime is expected to scale with the active stress 
$\alpha$ as $l_{\alpha} \sim \alpha^{-1/2}$~\cite{giomi2013defect}. By symmetry nematic activity is controlled by $v_0^2$. Assuming that the nematic stiffness is proportional to $J$, we 
estimate $\alpha\sim v_0^2 J$. We then estimate that the transition to turbulence will occur when we can accommodate $N$ defects of mean separation $\ell_\alpha$ on the surface of the sphere, 
i.e., $N=4\pi R^2/l_{\alpha}^2$. This gives a prediction for the transition to turbulence as 
\begin{equation} 
N \sim (R v_0)^2 J>4\;.
 \end{equation}
This form gives a very good  scaling collapse  of the observed number of defects for $J=10$ where there is no band state (Fig.~\ref{fig:Ndefects}b).  

\subsection{The bending state}

\begin{figure}[tb]
\centering
\includegraphics[width=0.9\columnwidth]{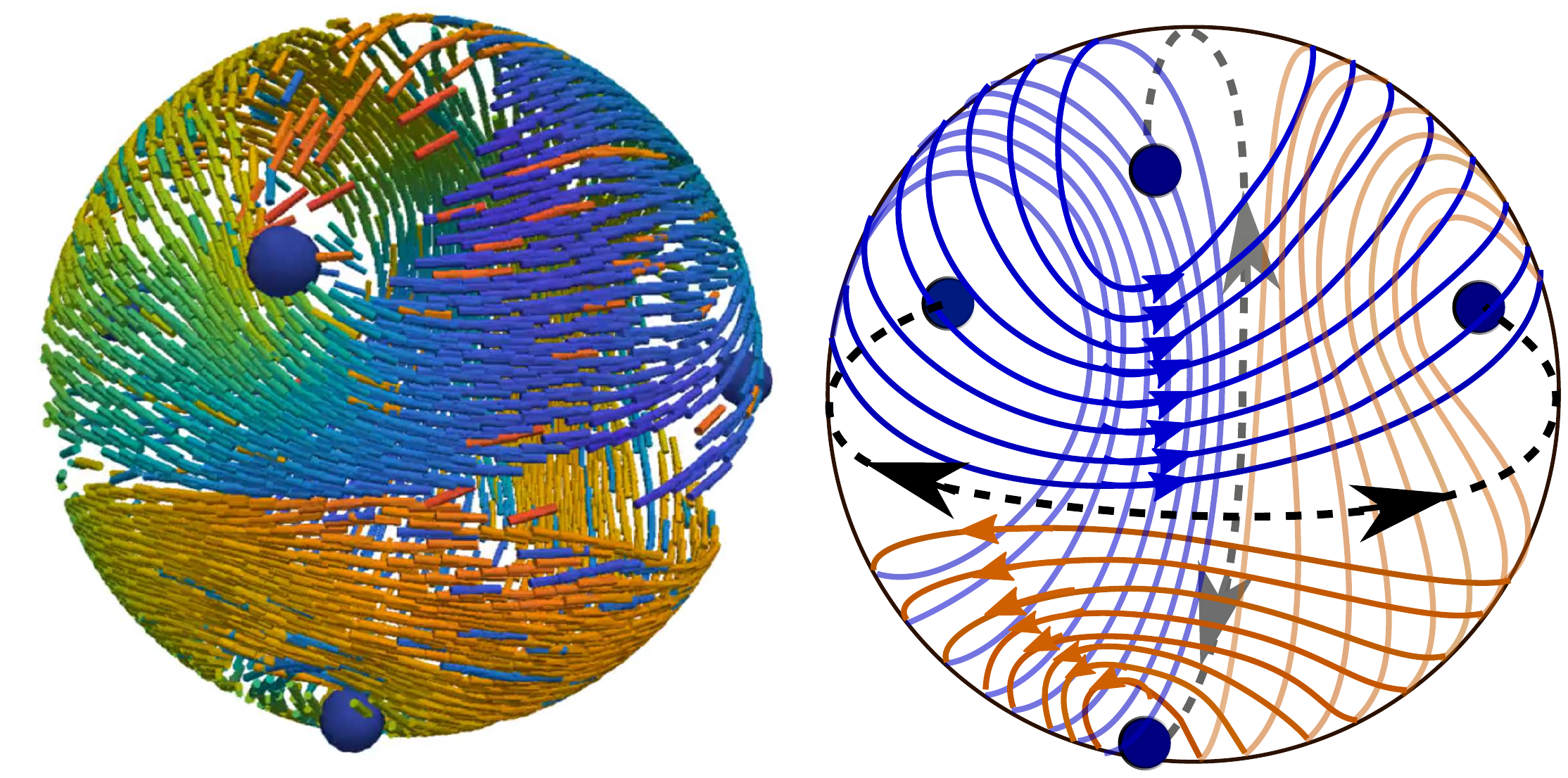}
\caption{Bending of the band at high activity and low alignment. The high persistence lengths favour polarity sorting, leading to one-directional polar flow. The $+1$ defects of the band 
split into pairs of $+1/2$ defects that are gradually pushed to the vertices of a tetrahedron. Left: Simulation snapshot at $J=0.1$, $v_0=3.0$. Right: Schematic drawing of the topological 
state.} 
\label{fig:Bending}
\end{figure}

At low values of $J$, as already noted, the transition to turbulence is delayed. Instead, we observe an intriguing state, where the band that forms at intermediate $v_0$ develop a bending  
instability (Fig.~\ref{fig:snapshots}d and Fig.~\ref{fig:Bending}). This is indeed seen in the experiments of Keber,\emph{et al.}~\cite{keber2014topology} and we have also observed it in the polar case \cite{inprep_topology}. It is tempting to speculate that on a sphere this ``bending of the bands'' may provide a generic route to the spatiotemporal chaos of the turbulent state, but a more quantitative 
analysis beyond the scope of the present work will be required to substantiate this idea.  Like in the experiment, the bending is intermittent: it is not seen in all runs for a
given set of parameters, and it can both appear and disappear in a given simulation. The instability occurs via the mechanism described pictorially in Fig.~\ref{fig:Bending}. It is energetically favourable 
for the  two $+1$ defects that sit at the poles in  the band state to split into four $+1/2$ defects~\cite{shin2008topological} because by doing so they can lower the total core energy of the configuration. 
At low $J$, the elastic energy for bending  a circulating band is low, and a bending instability can be triggered by the finite size of the sphere. This occurs when the persistence length of the nematic, 
$\xi=v_0\tau_{\mathrm{flip}}$, is comparable to the sphere circumference, so that agents can perform a full directed circulation around the sphere before reversing, resulting in lane formation and  
polar flows (see Fig.~S2 in \cite{SI}). The interplay between curvature, polar flow and steric effects then drives such an instability, similar to a Rossby wave or a garden hose instability~\cite{batchelor2000introduction}. The lowest mode of oscillation of the ``bent band'' at maximum amplitude is  compatible with the tetrahedral defect state of the nematic, as  the large meander of the band includes one $+1/2$ defect 
in every bend, resulting in a configuration with opposing streams touching at the front and the back of the sphere, as sketched out in Fig.~\ref{fig:Bending}. In Fig.~S5 in \cite{SI}, we show the evolution 
of the bending state with sphere radius $R$. Like in experiment, bending is observed only for sufficiently small systems where $v_0\tau_{\mathrm{flip}}\geq R$.

\section{Conclusions}
\label{sec:conclusions}

In this paper we presented results of detailed numerical simulations of a model of self-propelled agents with nematic alignment confined to move on the surface of a sphere. We developed algorithms 
for automatic detection and tracking of topological defects, which allowed us to characterise a series of emergent collective motion patterns and construct a detailed phase diagram. The onset of these 
patterns can be understood using ideas from liquid crystal theory, topology, and the study of active nematics in flat space. Our work provides a systematic framework for the experimental results of 
Keber, \emph{et al.}~\cite{keber2014topology}. Most of the states observed here have been seen in the experiment at different sphere radii $R$ and ATP concentrations. We believe that our phase 
diagram Fig.~\ref{fig:phase_diagram} will be especially useful. In contrast to the two existing hydrodynamic models \cite{keber2014topology,khoromskaia2016vortex}, we can access high-activity 
states with large density fluctuations, which are highly experimentally relevant as bands form there.
 
A full continuum analytical theory of this system is lacking at present. Taking the proper hydrodynamic limit of agent-based active nematic models at high densities is complex \cite{gao2015multiscale}. 
Continuum active nematic theories rely on the presence of extensile or contractile active stresses, but the origin of such stresses is unclear in particle-based models where activity is most naturally 
introduced as self-propulsion. In order to obtain nematic behaviour one has to introduce a flipping time scale $\tau_{\mathrm{flip}}$ of the director. As we showed here, the onset of most collective 
motion patterns is sensitive to this time scale. Furthermore, we showed that while the behaviour predicted by continuum theory emerges at large scales, there is an intermediate scale in which
the non-universal microscopic details of the model cannot be ignored and can actually dominate the physics. An important consequence is that defect motion in self-propelled active nematics, 
as shown here and in \cite{shi2013topological}, is very slow, only mildly superdiffusive, and dominated by local fluctuations. In contrast, in locally extensile systems such as~\cite{decamp2015orientational,alaimo2017curvature} a local flow field does seem to naturally emerge. Despite these limitations, we argue that agent-based models can provide valuable insight into 
the behaviour of active nematics.
  
One other hurdle to a direct comparison with experiment is that we have simulated disks, not polymers. Recent planar simulations of active
polymer melts by two of us show that hairpin bends in the filaments (which are also apparent in experiment) are strongly implicated in the defect dynamics~\cite{prathyusha2016dynamically}. 
In the future, for greater experimental relevance, we plan to directly simulate active polymers with an extensile activity mechanism. Preliminary results show that defect motion and oscillations 
dominate the motion, but only if all locally polar parts of the dynamics are suppressed. Finally, it would be desirable to move away from the \emph{dry} limit and explicitly include hydrodynamic effects.
 
\section{Acknowledgements}
The authors would like to acknowledge many valuable discussions with Mark Bowick, Daniel L.~Barton, and Prathyusha K.~R. This collaboration was made possible by a travel grant from the 
Northern Research Partnership (NRP) Fund. RS acknowledge support by UK BBRSC (grant BB/N009789/1) and SH acknowledges support by the UK BBSRC (grant BB/N009150/1). MCM was 
supported by the US National Science Foundation through awards DMR-1609208 and DGE-1068780 and by the Syracuse Soft Matter Program.

\appendix
\section{Identifying and tracking topological defects} 
\label{sec:appendix}
\begin{figure}[tb]
\centering
\includegraphics[width=0.99\columnwidth]{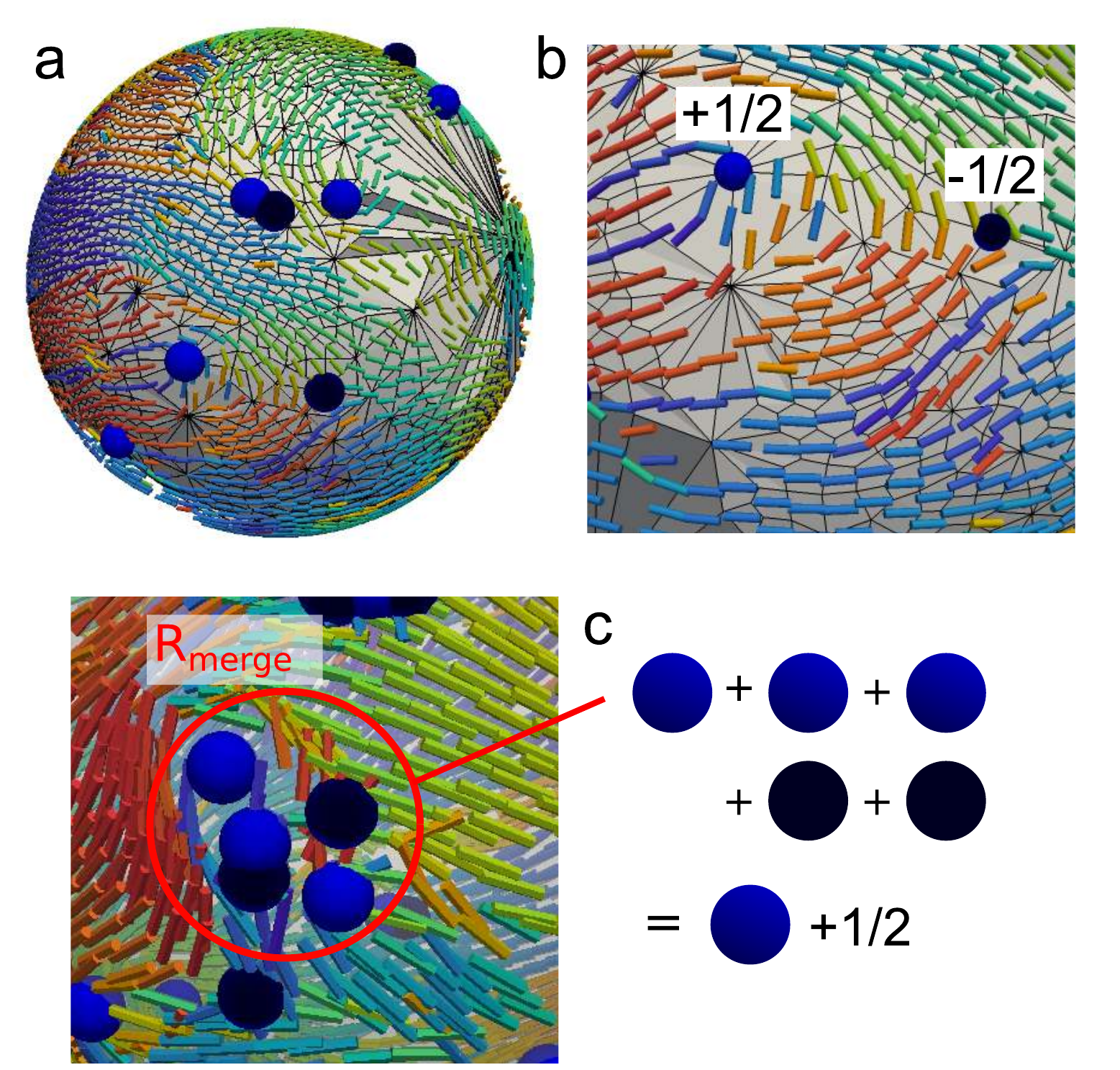}
\caption{(a) Example of the contact network and its dual (grey polygons) on a sphere with substantial density fluctuations. We use the same colour convention for the discrete director
field as in Fig.~\ref{fig:snapshots}. (b) Zoom-in onto a pair of a positive and a negative $1/2$ defects. (c) Illustration of the merging technique for spots with local disorder. The five 
$\pm 1/2$ defects within $R_{\text{merge}}$ combine into one $+1/2$ defect.} 
\label{fig:tesselation_main}
\end{figure}

We begin by discussing the algorithm used to identify and track topological defects. The identification of the defects is not completely straightforward due to large density 
fluctuations that accompany some of the collective motion patterns, where one has to distinguish between a defect and a bald spot. In order to discriminate between the two, we 
implemented a generalised computation of the winding number. For each agent we constructed a network of all of its neighbours within a cutoff distance, i.e.~the contact network. 
For systems with large density fluctuations it is not practical to impose a predefined cutoff, and instead we iteratively adjusted the cutoff distance until each agent had at least three 
neighbours. We then used the contact network to construct a polygonal tessellation of the sphere. Finally, we connected geometric centres of each of those polygons to construct the 
dual lattice. Compared to, e.g., a Voronoi diagram and its dual Delaunay triangulation, our method has the advantage that it generalises to any orientable surface. As a result of this 
construction, each vertex of the dual lattice is in the centre of a polygonal loop with the nematic director, $\mathbf{n}$, or velocity, $\mathbf{v}$, assigned to each of its corners. Each 
loop then acts as a discrete integration path for $\mathbf{n}$ or $\mathbf{v}$. Finally, we projected the field vectors onto the local tangent plane (plane of the polygonal loop), and 
integrated angle differences between two consecutive vertices in the counterclockwise direction around the contour, giving us the half-integer values of the topological charges for 
each loop centre. In Fig.~\ref{fig:tesselation_main}a-b we show examples of $\mathbf{n}$-fields, tessellations and defects. Defects together with their charges are shown in 
Fig.~\ref{fig:snapshots} and Fig.~\ref{fig:Bending}. Determining the total number of defects allows us to easily identify the transition to turbulence (Fig. \ref{fig:Ndefects}). 

We then tracked defects between consecutive snapshots of a simulation by matching defects of a given charge with the nearest defect with the same charge in the previous frame.
To reduce spurious tracking errors due to proliferation of defect clusters at spots with local disorder, we first merges all defect charges within a radius $R_{\text{merge}}=5\sigma$, as shown in 
Fig.~\ref{fig:tesselation_main}c. We were able to track $+1/2$ defects in the four defect, merging and bending states, and $+1$ defects in the band state. We did not attempt to track $-1/2$ defects,
or defects in the turbulent state. Using spherical coordinate system, position of each defect was identified by a unique $\theta$ and $\phi$. Fig.~\ref{fig:snapshots} shows the stereographic 
projections of  representative defect trajectories for the various states with the defects tracked in different
colours. The corresponding trajectories in the sphere are shown in Fig S2 of \cite{SI}.  
\vspace{-2mm}

\clearpage


\setcounter{equation}{0}
\setcounter{figure}{0}
\setcounter{table}{0}
\makeatletter
\renewcommand{\theequation}{S\arabic{equation}}
\renewcommand{\thefigure}{S\arabic{figure}}
\renewcommand{\bibnumfmt}[1]{[S#1]}
\renewcommand{\citenumfont}[1]{S#1}

\pagebreak
\widetext
\vspace{5mm}
\begin{center}
	\textbf{\large Dynamical patterns in active nematics on a sphere: Supplementary information}
\end{center}
\vspace{15mm}
\section*{Band profiles}
\begin{figure*}[h]
\centering
\includegraphics[width=0.99\textwidth]{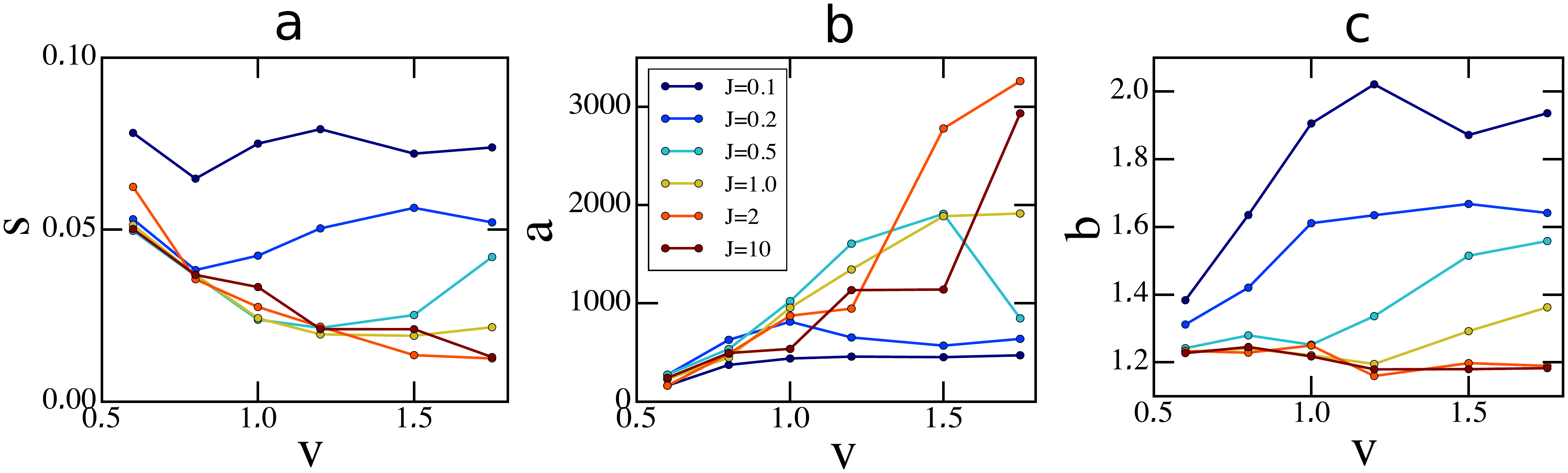}
\caption{Results of the fits of the angle and density profiles to the polar band theory.}
\label{fig:bandfits_supplementary}
\end{figure*}

In Ref. [23] of the main text, we derived an effective equation for the band density 
\begin{align}
& \rho(\theta)=a\left[(\sin s\theta\!-\!\sin s\theta_m)\right]+b,\quad \theta \in [\theta_m, \pi - \theta_m] \label{eq:profile} \\
& \rho(\theta)=0 \quad \text{otherwise,} \nonumber
\end{align}
where $a=\frac{1}{2R} \left[\frac{R}{2\sigma}\right] \frac{v_0}{\mu k s}$ and $b/a=\frac{2\sigma}{R} s \cos s \theta_m + \frac{1}{a}$ for the polar case (note that in [23] , 
we chose $\theta=0$ at the equator). Eq.~\eqref{eq:profile} describes a sine profile symmetric about the equator between two band edges located at $\theta_m$ and $\pi-\theta_m$.    
Here $s$ is the effective coupling strength between the active force direction $\mathbf{n}$ and the angle from the equator. It can be measured as the slope of the $\mathbf{n}$-angle 
profiles $\alpha(\theta)$. An example of such a profile in the full band state at $J=0.1$  is in the inset of Fig. 6a of the main text. The slope is clearly visible, though much shallower than in the polar case. The large deviations are in the edge regions where the density drops to zero.

In Figure \ref{fig:bandfits_supplementary}, we fit equation \ref{eq:profile} to the nematic bands.  In panel a, we show the $s$ from fits to the slope in the middle parts of the profiles. Like in the polar case, $s$ decreases with $J$ from a maximum of $s=0.07$ at $J=0.1$ until it reaches a plateau at $s\approx 0.03$ for $J\geq0.5$. This is consistent with the scaling of the density profiles shown in Fig. 6b-d of the main text. The low values of $s$ are also consistent with the delayed appearance of the band state in the nematic system compared to the polar one. We also notice that unlike for the polar case, for the nematic case $s$ depends on $v_0$: After an initial peak that coincides with the first appearance of the bands, $s$ decreases until it hits a $J$-dependent plateau value. Finally, in panel b and c we show the values of $a$  and $b$ extracted from fits to eq. \ref{eq:profile} using the $s$ from panel a. Consistent with the polar case and the increasingly peaked density profiles that we observe, $a$ increases with $v_0$, though not in the linear fashion of the analytical prediction. 

\begin{figure*}[h!]
\centering
\includegraphics[width=0.85\textwidth]{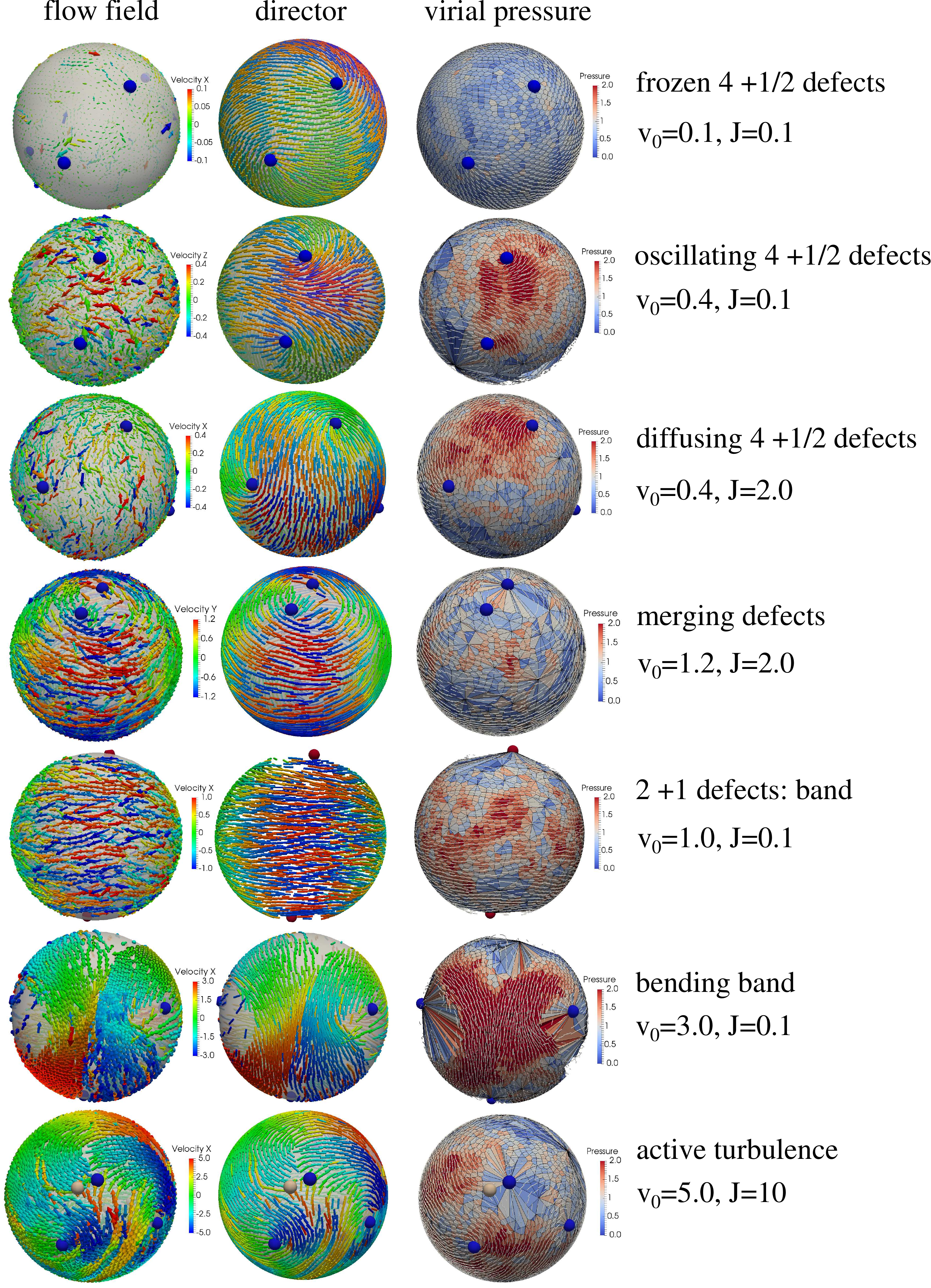}
\caption{Flow fields (left), nematic director fields (middle) and pressure (right) in the different dynamical patterns that we observe. Here both flow field and nematic field have been colored by the same component of motion as indicated in the caption for the velocity. The limits on the velocity color bars correspond to $(-v_0,v_0)$. For the virial pressure, we compute $p_i = \sum_j \mathbf{r}_{ij} \cdot \mathbf{F}_{ij} / A_i$ for each agent.  Here the sum runs over the contact neighbors $j$ of $i$, and $A_i$ is the area of the tile belonging to $i$ computed from the dual to the contact network. On the image, each tile has been colored by its pressure. As in Fig. 1 of the main text, $+1/2$ defects are blue, $-1/2$ defects are tan, and $+1$ defects are red.}
\label{fig:flow_fields_supplementary}
\end{figure*}

\begin{figure*}[t]
\centering
\includegraphics[width=1.0\textwidth]{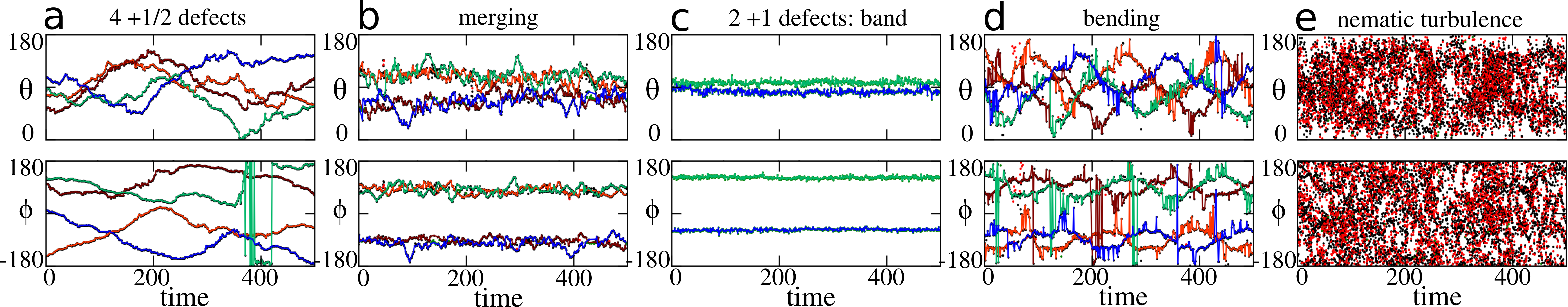}
\caption{Defect tracks as a function of time for the same simulation runs as the stereographic projections in Fig. 1 of the main text. a: Four defects for $v_0=0.4$ and $J=1$, b: merging defects for $v_0=1.2$ and $J=2$, c: band state for $v_0=1.2$ and $J=0.1$, d: bending bands for $v_0=2.5$ and $J=0.1$, e: active turbulence for $v_0=5$ and $J=5$. }
\label{fig:stereographic_supplementary}
\end{figure*}

\begin{figure*}[t]
\centering
\includegraphics[width=1.0\textwidth]{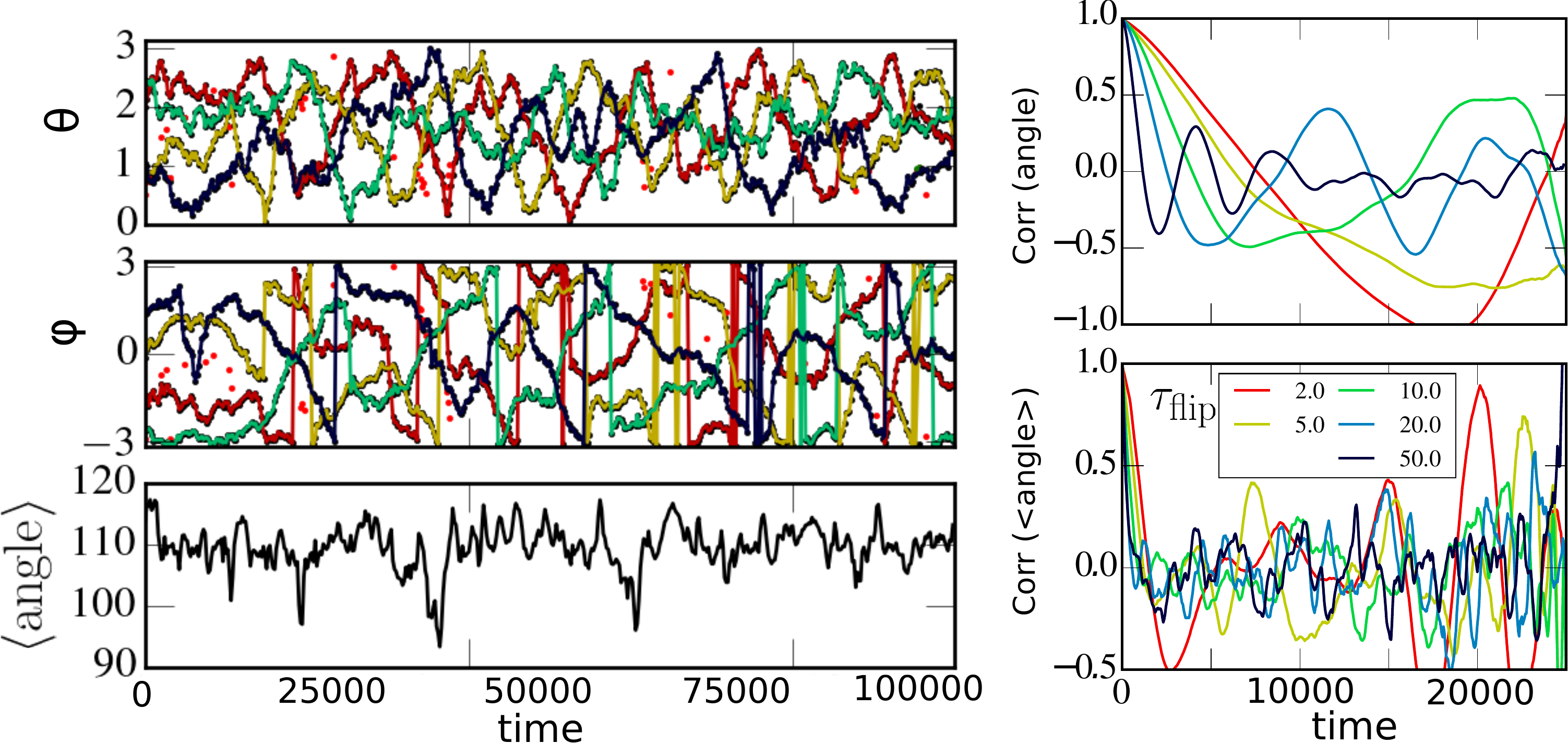}
\caption{Oscillations at larger flip times $\tau_{\text{flip}}=2-50$ for $v=0.4$ and $J=0.1$ compared to Fig. 5 of the main text. In the left three panels, we show the defect trajectories in polar coordinates (top two panels), and the mean angle $\langle \theta_{ij} \rangle$ (bottom panel) for $\tau=50$. The right two panels show the defect trajectory autocorrelations (top) and the mean angle autocorrelations (bottom). The oscillations frequency increases with $\tau_{\text{flip}}$ in the range shown.} 
\label{fig:oscillations_supplementary}
\end{figure*}

\begin{figure*}[t]
\centering
\includegraphics[width=1.0\textwidth]{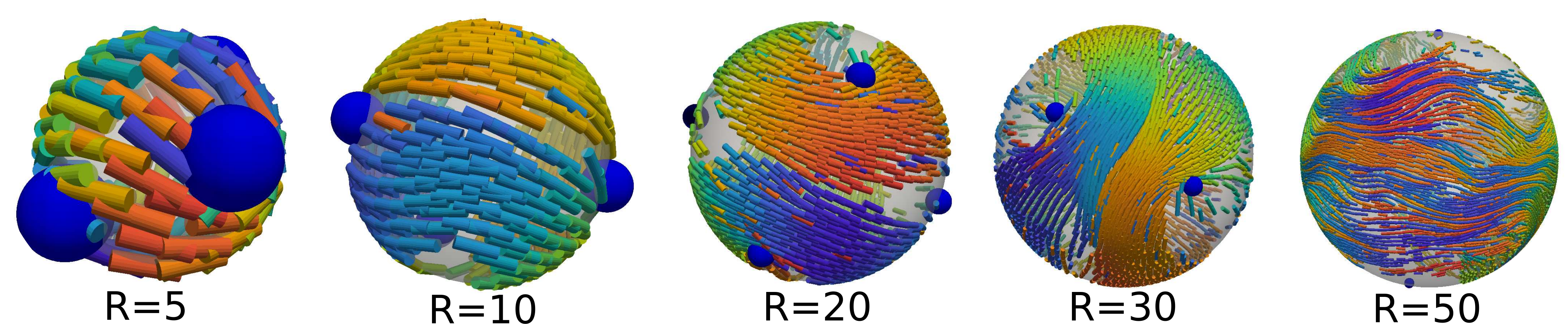}
\caption{Bending states as a function of system size. This is an intermittent state which appears for only some of the simulation runs / part of a simulation run. For $\tau_{\text{flip}}=100$, the bending state disappears for $R\geq 50$. All systems are at $J=0.1$, and we have $v_0=1.0$ for $R=5$, and $v_0=3.0$ for $R=10-50$.}
\label{fig:bending_supplementary}
\end{figure*}

\end{document}